\documentclass[conference]{IEEEtran}

\usepackage[letterpaper, left=1.02in, right=1.02in, bottom=1in, top=0.75in]{geometry}
\usepackage{graphicx}
\usepackage{amsmath}
\usepackage{amssymb}
\usepackage{float}
\usepackage{mathtools}
\usepackage{algorithmic}
\usepackage{algorithm}

\usepackage{amsmath}
\usepackage{multirow}
\usepackage{amsthm}

\usepackage{tablefootnote}
\usepackage{threeparttable}

\newcommand\floor[1]{\lfloor#1\rfloor}
\newcommand\ceil[1]{\lceil#1\rceil}
\newcommand{\nid}{\noindent}

\newcommand{\argmin}{\operatornamewithlimits{argmin}}

\IEEEoverridecommandlockouts
\DeclareMathOperator{\sgn}{sgn}

\begin{document}
%
\title{Uniquely Decodable Ternary Codes for\\Synchronous CDMA Systems\thanks{This work was partially supported by Natural Sciences and Engineering Research Council (NSERC).}}

\author{\IEEEauthorblockN{Michel Kulhandjian}
	\IEEEauthorblockA{School of Electrical Engineering\\ and Computer Science\\
		University of Ottawa\\
		Ottawa, Ontario, K1N 6N5, Canada\\
		E-mail: mkk6@buffalo.edu}
	\and
	\IEEEauthorblockN{Claude D'Amours}
	\IEEEauthorblockA{School of Electrical Engineering\\ and Computer Science\\
		University of Ottawa\\
		Ottawa, Ontario, K1N 6N5, Canada\\
		E-mail: cdamours@uottawa.ca}
	\and
	\IEEEauthorblockN{Hovannes Kulhandjian}
	\IEEEauthorblockA{Department of Electrical\\ and Computer Engineering\\
		California State University, Fresno\\
		Fresno, CA 93740, U.S.A.\\
		E-mail: hkulhandjian@csufresno.edu}}


%



\maketitle

\begin{abstract}
In this paper, we consider the problem of recursively designing uniquely decodable ternary code sets for highly overloaded synchronous code-division multiple-access (CDMA) systems. The proposed code set achieves larger number of users $K < K_{max}^t$ than any other known state-of-the-art ternary codes that offer low-complexity decoders in the noisy transmission. Moreover, we propose a simple decoder that uses only a few comparisons and can allow the user to uniquely recover the information bits. Compared to maximum likelihood (ML) decoder, which has a high computational complexity for even moderate code length, the proposed decoder has much lower computational complexity. We also derived the computational complexity of the proposed recursive decoder analytically. Simulation results show that the performance of the proposed decoder is almost as good as the ML decoder. 
\end{abstract}


%
\IEEEpeerreviewmaketitle
\vspace{-0 mm}

\section{Introduction}\vspace{-0 mm}
The uniquely decodable coding methods for overloaded synchronous code-division multiple-access (CDMA) where the number of multiplexed signals $K$ is greater than the spreading gain or code (signature) length $L$ has been studied in \cite{Shapiro1963}-\cite{Marvasti2016}. An overloaded code set $\mathbf{C}$ of dimension $L\times K$ is considered to be ``\textit{errorless}'', or uniquely decodable (UD) in a noiseless multiplexed transmission if for all possible $K\times 1$ vectors $\mathbf{x}_1$ and $\mathbf{x}_2$, where $\mathbf{x}_1 \neq \mathbf{x}_2 \in \{\pm1\}^{K \times 1}$ and $\mathbf{C}\mathbf{x}_1 \neq \mathbf{C}\mathbf{x}_2$ \cite{Ferguson1982}. In other words, a UD matrix is injective in nature or there exists a one-to-one mapping between the input and output. 

Uniquely decodable overloaded code set construction for noiseless channel where $f(L)$ represents the maximum number of columns (signals) that matrix can have for a given $L$ and still be uniquely decodable is related to coin-weighing problem, one of the Erd\"{o}s's problem in \cite{Erdos1963}. In the literature the explicit construction techniques of binary $(0,1)$, antipodal $(\pm 1)$, and ternary $(0, \pm 1)$ have been investigated in \cite{Shapiro1963}, \cite{Lindstrom1964}-\cite{Khachatrian1989}, \cite{Khachatrian1987}-\cite{michel2012}, and \cite{Cantor1966}, \cite{Chang1979}-\cite{Khachatrian1998}, \cite{Marvasti2012}-\cite{Marvasti2016} most of which are recursive in nature. To the best of our knowledge, the maximum number of vectors of the explicit constructions of binary, antipodal and ternary code sets are $K_{max}^b = \gamma(L+1)\footnote{where $\gamma(n)$ function is the number of ones in the binary expansion of all positive integers less than n.}$, $K_{max}^a = \gamma(L) +1$ and $K_{max}^t = (k+2)2^{(k-1)}$,
as shown in Table \ref{table:binary}, Table \ref{table:antipodal} and Table \ref{table:ternary}, respectively.
\begin{table*}[h]
	\caption{Binary Codes} 
	\centering 
	\begin{threeparttable}
		\begin{tabular}{l l c c c c} 
			\hline\hline  
			\multicolumn{1}{c}{\multirow{2}{*}[-1.5pt]{\bf{Year}}} &
			\multicolumn{1}{l}{\multirow{2}{*}[-1.5pt]{\bf{Authors and Publications}}}  & \multicolumn{1}{c}{\multirow{2}{*}[-1.5pt]{$\bf{n}$}} & \multicolumn{1}{c}{\multirow{2}{*}[-1.5pt]{$\bf{K}$}} & \multicolumn{2}{c}{\multirow{1}{*}[-1.5pt]{\bf{Decoder}}} \\[0.5ex]  \cline{5-6}
			&  &  &  & \multirow{1}{*}[-1.5pt]{\bfseries{Noiseless}} & \multirow{1}{*}[-1.5pt]{\bfseries{AWGN}}\\ [1.0ex]
			\hline   \rule{-3pt}{2.5ex} 
			1963 & S\"{o}derberg and Shapiro \cite{Shapiro1963} & $L$  & $<\gamma(L+1)$ & No  & No\\[0.6ex]
			1964 & Lindstr\"{o}m  \cite{Lindstrom1964}  & $L$ & $\bf{\boldsymbol{\gamma}(L+1)}$\tnote{\dag} & No & No \\[0.6ex]
			1966 & Cantor and Mills \cite{Cantor1966}  & $2^k-1$ & $\bf{k2^{(k-1)}}$ & No & No \\[0.6ex]
			1989 & Martirossian and Khachatrian \cite{Khachatrian1989}  & $L$ & $\bf{\boldsymbol{\gamma}(L+1)}$ & Yes & No \\[0.6ex]
			\hline 
		\end{tabular}
		\footnotesize
		\begin{tablenotes}
			\item[\dag] Code set constructions that achieve the maximum number of vectors $\mathbf{K}_{max}$ are presented in bold.
		\end{tablenotes}
	\end{threeparttable}
	
	\label{table:binary}
\end{table*}

\begin{table*}[h]
	\caption{Antipodal Codes} 
	\centering 
	\begin{tabular}{l l c c c c} 
		\hline\hline  
		\multicolumn{1}{c}{\multirow{2}{*}[-1.5pt]{\bf{Year}}} &
		\multicolumn{1}{l}{\multirow{2}{*}[-1.5pt]{\bf{Authors and Publications}}}  & \multicolumn{1}{c}{\multirow{2}{*}[-1.5pt]{$\bf{n}$}} & \multicolumn{1}{c}{\multirow{2}{*}[-1.5pt]{$\bf{K}$}} & \multicolumn{2}{c}{\multirow{1}{*}[-1.5pt]{\bf{Decoder}}} \\[0.5ex]  \cline{5-6}
		&  &  &  & \multirow{1}{*}[-1.5pt]{\bfseries{Noiseless}} & \multirow{1}{*}[-1.5pt]{\bfseries{AWGN}}\\ [1.0ex]
		\hline   \rule{-3pt}{2.5ex} 
		1964 & Lindstr\"{o}m \cite{Lindstrom1964} & $L$  & $\bf{\boldsymbol{\gamma}(L)+1}$ & No  & No\\[0.6ex]
		1987 & Khachatrian and Martirossian \cite{Khachatrian1987}  & $L$ & $\bf{\boldsymbol{\gamma}(L)+1}$ & No & No \\[0.6ex]
		1995 & Khachatrian and Martirossian \cite{Khachatrian1995}  & $2^k$ & $\bf{k2^{(k-1)}+1}$ & Yes & No \\[0.6ex]
		2012 & Kulhandjian and Pados \cite{michel2012}  & $2^k$ & $\bf{k2^{(k-1)}+1}$ & Yes & No \\[0.6ex]
		\hline 
	\end{tabular}
	\label{table:antipodal}
\end{table*}

\begin{table*}[h]
	\caption{Ternary Codes} 
	\centering 
	\begin{tabular}{l l c c c c} 
		\hline\hline  
		\multicolumn{1}{c}{\multirow{2}{*}[-1.5pt]{\bf{Year}}} &
		\multicolumn{1}{l}{\multirow{2}{*}[-1.5pt]{\bf{Authors and Publications}}}  & \multicolumn{1}{c}{\multirow{2}{*}[-1.5pt]{$\bf{n}$}} & \multicolumn{1}{c}{\multirow{2}{*}[-1.5pt]{$\bf{K}$}} & \multicolumn{2}{c}{\multirow{1}{*}[-1.5pt]{\bf{Decoder}}} \\[0.5ex]  \cline{5-6}
		&  &  &  & \multirow{1}{*}[-1.5pt]{\bfseries{Noiseless}} & \multirow{1}{*}[-1.5pt]{\bfseries{AWGN}}\\ [1.0ex]
		\hline   \rule{-3pt}{2.5ex} 
		1966 & Cantor and Mills \cite{Cantor1966} & $2^k$  & $\bf{(k+2)2^{(k-1)}}$ & No  & No\\[0.6ex]
		1979 & Chang and Weldon \cite{Chang1979}  & $2^k$ & $\bf{(k+2)2^{(k-1)}}$ & Yes & No \\[0.6ex]
		1982 & Ferguson \cite{Ferguson1982}   & $2^k$ & $\bf{(k+2)2^{(k-1)}}$ & Yes & No \\[0.6ex]
		1984 & Chang \cite{Chang1984}  & $2^k$ & $\bf{(k+2)2^{(k-1)}}$ & No & No \\[0.6ex]
		1998 & Khachatrian and Martirossian \cite{Khachatrian1998}  & $2^k$ & $\bf{(k+2)2^{(k-1)}}$ & Yes & No \\[0.6ex]
		2012 & Mashayekhi and Marvasti \cite{Marvasti2012}  & $2^k$ & $2^{(k+1)}-1$ & Yes & Yes \\[0.6ex]
		2016 & Singh \textit{et al.} \cite{Marvasti2016}   & $2^k$ & $2^{(k+1)}-2$ & Yes & Yes \\[0.6ex]
		2018 & Proposed   & $2^k$ & $2^{(k+1)}+2^{(k-2)}-1$ & Yes & Yes \\[0.6ex]
		\hline 
	\end{tabular}
	\label{table:ternary}
\end{table*}
Those code sets, which are primarily designed for the noiseless channel, have relatively fast, very low complexity, recursive deterministic decoders. In noisy channels one may apply the optimal decoder such as maximum likelihood (ML); however, the computational complexity grows with the code length and it is not very practical. Recently, in \cite{ming2016}, a class of antipodal code sequences, which hierarchically possess cross-correlation, for overloaded code-division multiplexing (CDM) systems with simplified two-stage ML detection has been proposed. In addition to that other overloaded matrices over the ternary alphabet are introduced in \cite{Marvasti2012} with fast logical decoder, which requires few comparisons. Similarly, in \cite{Marvasti2016} the authors propose overloaded code sets over the ternary alphabet that have twin tree structured cross-correlation hierarchy with a simple multi-stage detection. One potentially can take advantage of such codes' structure and decoding scheme and make use in non-orthogonal multiple access (NOMA) schemes that recently have received significant attention for the fifth generation (5G) cellular networks \cite{michel2017}.

In this work, we consider the problem of recursive uniquely decodable ternary code construction method for highly overloaded synchronous CDMA systems. Although the overloaded factor  $ \frac{K}{L}$ increases in the sequence of code set they remain uniquely decodable. The proposed decoder is designed in a such a way that the user can uniquely recover the information bits with a very simple decoder, which uses only a few comparisons. In contrast to ML decoder, the proposed decoder has much lower computational complexity. Simulation results in terms of bit error rate (BER) demonstrate that the performance of the proposed decoder is very close to that of the ML decoder.    
\\
\\
The rest of the paper is organized as follows. In Section \ref{construction}, we present the construction of the uniquely decodable code sets followed by the decoding algorithm in Section \ref{decoder}. In Section \ref{performanceAnalysis}, the complexity of the proposed code set's decoding scheme is analyzed. In Section \ref{simulation}, we present our simulation methodology and results before presenting our conclusions in Section \ref{conclusion}.

The following notations are used in this paper. All boldface lower
case letters indicate column vectors and upper case letters indicate
matrices, $()^T$ denotes transpose operation,  $\mathbb{C}$ denotes
the set of all complex numbers, $\mod$ denotes the modulo operation, $rnd$ stands for round to the nearest integer function, $\sgn$ denotes the sign function, $|\cdot|$ denotes complex amplitude, $\ceil{.}$ is the ceiling function and $\floor{.}$ is the floor function, respectively.

\section{{Recursive Code Construction}}
\label{construction}

We recall that a ternary code set $\mathbf{C} \in \{0,\pm 1\}^{L \times K}$ is uniquely decodable over signals $\mathbf{x} \in \{ \pm 1 \}^{K\times 1}$ or $\mathbf{x} \in \{ 0, 1 \}^{K\times 1}$, $K > L$, if and only if, for any $\mathbf{x}_1 \neq \mathbf{x}_2$, $\mathbf{C}\mathbf{x}_1 \neq \mathbf{C}\mathbf{x}_2$ or, equivalently, $\mathbf{C}(\mathbf{x}_1 - \mathbf{x}_2)\neq \mathbf{0}_{L \times 1} $. We can rewrite the unique decodability necessary and sufficient condition as $\mathsf{Null}(\mathbf{C}) \cap \{0,\pm 2\}^{K\times 1} = \{ 0\}^{K\times 1}$ or in an equivalent manner as
\begin{equation}
\label{null01}
\mathsf{Null}(\mathbf{C}) \cap \{0,\pm 1\}^{K\times 1} = \{ 0\}^{K\times 1}.
\end{equation}
Let $f_t (L)$ represent the maximum number of columns (signals) that matrix can have for a given $L$ and still be
uniquely decodable. For the ternary code matrix with codes of length $L=2$, $f_t(2)$ is simple and can be found by looking at the total number of possible columns $3^2 = 9$. Excluding the $[0,0]^T$ column, half of the remaining is the negative of the other half, which makes it a total of $4$ distinct columns that can be chosen to be $[0,1]^T$, $[1,0]^T$, $[1,-1]^T$, and $[1,1]^T$. We conclude that no possible distinct combinations of these $4$ columns satisfy uniquely decodability criteria (\ref{null01}). Out of all the possible combinations there are only few matrices with number of columns of $3$ that satisfy (\ref{null01}), therefore $f_t(2)=3$. Every possible matrix of dimension $2\times 3$ that has uniquely decodable property can be reduced to\begin{equation}
\mathbf{C}_{2\times3}^1 =\begin{bmatrix*}[r]
+1 & +1  & +1 \\
+1 & 0  & -1 
\label{matrix1}
\end{bmatrix*},
\end{equation}
\nid by applying operations such as multiplying columns by negative one, permuting rows and columns.

For the case of $L=3$ and $L=4$ it can be shown with an exhaustive search that $f_t(3) = 5$ and $f_t(4) = 8$, respectively. In the preparation of general construction of matrices having $L=2^i$, where $i\geq2$, we carefully choose our seed matrix $\mathbf{C}_{4\times8}^2$ from distinct uniquely decodable matrices, which are found by exhaustive search, 
\begin{equation}
\mathbf{C}_{4\times8}^2 =\begin{bmatrix*}[r]
+1 & +1 & +1 & +1 & +1 & +1 & +1 & +1\\
+1 & +1 & +1 & +1 & 0 & -1 & -1 & -1\\
+1 & +1 & 0 & -1 & 0 & +1 & 0 & -1\\
+1 & 0 & 0 & -1 & 0 & -1 & 0 & +1
\label{matrix2}
\end{bmatrix*}.
\end{equation}
Now, we are ready to propose a general $L_i\times K_i$ code set design for $L_i=2^i$ with $K_i=2^{i+1}+2^{i-2}-1$, $i=3,4,...\;$. Starting from  $\mathbf{C}_{4\times8}^2$ the following recursive relation defines a sequence of matrices. The $i^{th}$ recursive matrix  $\mathbf{C}_{L_i\times K_i}^i$ is formed as follows:
\begin{equation}
\mathbf{C}_{L_i\times K_i}^i =\begin{bmatrix*}[r]
+1 & \dots & +1 & +1 & +1   & \dots & +1 \\
+1 & \dots & +1 &  0 &  -1  & \dots & -1 \\
&     &    &  0 &      &     &  \\
& \mathbf{\hat{C}}^{i-1} &  & 0 &  & \mathbf{0} &  \\
&     &    &  \vdots &      &     &  \\
&   \mathbf{0}  &    &  0 &      &  \mathbf{\hat{C}}^{i-1}   &  \\
&     &    &  0 &      &     &
\label{matrixi}
\end{bmatrix*},
\end{equation}
where $L_i = 2 L_{i-1}$, $K_i = 2 K_{i-1} + 1$, $\mathbf{\hat{C}}^{i-1}$ is derived by eliminating the first row of $\mathbf{C}_{L_{i-1}\times K_{i-1}}^{i-1}$. We need to show that code sequences $\mathbf{C}_{L_i\times K_i}^i$ preserve the uniquely decodability property. Based on the assumption on $i$, assume that $\mathbf{C}^{i-1}$ is uniquely decodable and $\mathbf{y} = \mathbf{C}_{L_i\times K_i}^i \mathbf{x}$, where $\mathbf{x} \in \{\pm 1\}^{{K_i\times 1}}$. By looking at the first element of $\mathbf{y}$, $y_1 \in \{\pm K_i, \pm (K_i-2), ..., \pm1\}$, we can definitely find the number of $-1$'s in $\mathbf{x}$ to be $n = (K_i - y_1)/2 $. Considering the same argument, having the knowledge of $n$ combined with $y_2 \in \{\pm (K_i-1), \pm (K_i-3), ..., 0\}$, the number of $-1$'s in the first and last $K_{i-1}$ elements of $\mathbf{x}$, $n_l$ and $n_r$ can be uniquely determined to be $n_l=\floor{(2n-y_2)/4}$, $n_r = \floor{n-n_l}$. Note that if $n = n_l + n_r +1$ the middle element of $\mathbf{x}$ is $-1$ else it is $+1$. Since there is one-to-one mapping between $(y_1,y_2) \rightarrow (n, n_l, n_r)$ values accompanied with $\mathbf{\hat{C}}^{i-1}$ it can be shown that $[y_3,\dots,y_{L_i}]$ is uniquely generated by the first and last $K_{i-1}$ elements of $\mathbf{x}$. Therefore, we can conclude that $\mathbf{C}^i$ are uniquely decodable code sequences. 

\section{The Proposed Fast Decoder}
\label{decoder}
In the overloaded (i.e., $K>L$) synchronous code-division multiple-access application of interest, each user multiplexes its antipodal data, ($\pm 1$), using binary-phase shift keying (BPSK), by multiplying it with the signature and then transmitting it through the channel after carrier modulation. In a system with signature matrix $\mathbf{C}\in \{\pm 1, 0\}^{L \times K}$ in which the columns are the user vectors (spreading codes), the received vector can be expressed by
\begin{eqnarray}
\mathbf{y} &=& A\mathbf{C}\mathbf{x} + \mathbf{n} \\
&=& \sum_{j=1}^K A\mathbf{c}_j x_j + \mathbf{n}
\end{eqnarray}
\noindent where $A$ is the amplitude, $\mathbf{c}_j \in \{\pm 1, 0\}^{L \times 1}$ are signatures for $1\leq j \leq K$, $\mathbf{x} \in \{\pm 1\}^{K\times 1}$ is user data and $\mathbf{n}$ is additive white Gaussian noise (AWGN) channel noise. 

The objective of the receiver is the following; given the received vector $\mathbf{y}$ and $\mathbf{C}$ recover the user data $\mathbf{\hat{x}}$ such that the mean square error $E\{||\mathbf{x}-\mathbf{\hat{x}}||^2\} $ is minimized. It is known that obtaining the ML solution is generally NP-hard \cite{Lupas1989}. 

For our detection problem, where the overloaded signature matrix has UD structure, can be solved efficiently if there is a function that maps $\mathbf{y} \mapsto \widehat{\mathbf{y}} \in \Lambda$, where $\Lambda$ is a $\mathbb{Z}$-module with rank $L$. It is equivalent to finding the closest point in a lattice $\Lambda$, such that
\begin{eqnarray}
\label{MinDistY}
\widehat{\mathbf{y}} = \argmin_{\mathbf{y}' \in \Lambda} || \mathbf{y}- \mathbf{y}'||^2.
\end{eqnarray}

Gaining the knowledge of $\widehat{\mathbf{y}}$, one of the points in $\Lambda$ generated by $\mathbf{C}$, we can obtain  $\mathbf{\hat{x}}$ uniquely, since $\mathbf{C}$ satisfies the uniquely decodability criteria (\ref{null01}). However, there is no known polynomial algorithm that can obtain $\widehat{\mathbf{y}}$ from $\mathbf{y}$.

Therefore, we present the general form of the proposed fast decoding algorithm (FDA) for the $\mathbf{C}_{L_i\times K_i}^i$, $i\geq2$ case.
\vspace{-0.3cm}
\begin{center}
	\begin{table}[h]
		
		\begin{center}
			\begin{tabular}{l}
				\hline \hline \rule{0pt}{3ex} 
				\nid \textbf{Fast Decoder Algorithm (FDA)}  \\
				\hline \rule{0pt}{3ex} 
				\nid \textbf{{Input}:} $\mathbf{y}$ \\
				\hspace{0.3cm} 1: $z_1 \gets Q(y_1, -K,K)$ \\
				\hspace{0.3cm} 2: \textbf{If} $|z_1| = K$,   $\mathbf{\hat{x}}\gets sgn(z_1)\mathbf{1}$\\ 
				\hspace{0.3cm} 3:  \textbf{else}\\
				\hspace{0.3cm} 4:  \hspace{0.3cm} $n \gets (K-z_1)/2$\\
				\hspace{0.3cm} 5:  \hspace{0.3cm} $z_2 \gets Q(y_2, -(K-|z_1|),K-|z_1|)$\\
				\hspace{0.3cm} 6:  \hspace{0.3cm} $n_l\gets (2n-z_2)/4$, $n_r \gets n-n_l$ \\
				\hspace{0.3cm} 7:  \hspace{0.3cm} $n_l \gets \floor{n_l}$, $n_r \gets \floor{n_r}$ \\
				\hspace{0.3cm} 8:  \hspace{0.3cm}  \textbf{If} $K == 8$,   $\mathbf{\hat{x}} \gets subDecoder(\mathbf{y}, n_l,n_r)$ \\
				\hspace{0.3cm} 9:  \hspace{0.3cm}  \textbf{else} \\
				\hspace{0.3cm}10:  \hspace{0.6cm} $\mathbf{\hat{y}}_l\!\! \gets\!\! [(2^i+2^{i-3} - 1 - 2n_l), y_3, \dots, y_{2^{i-1}+1}]^T$ \\
				\hspace{0.3cm}11:  \hspace{0.6cm} $\mathbf{\hat{y}}_r \!\!\gets\!\! [(2^i+2^{i-3} - 1 - 2n_r), y_{2^{i-1}+2},\dots, y_{2^i}]^T$ \\
				\hspace{0.3cm}12:  \hspace{0.6cm} $\mathbf{\hat{x}}_l \!\!\gets\!\! decoder( \mathbf{\hat{y}}_l)$, $\mathbf{\hat{x}}_r \gets decoder( \mathbf{\hat{y}}_r)$  \\
				\hspace{0.3cm}13:  \hspace{0.6cm} $x_m \gets z_l -(\mathbf{\hat{x}}_l^T\mathbf{1} +\mathbf{\hat{x}}_r^T\mathbf{1}) $, $\mathbf{\hat{x}} \gets [\mathbf{\hat{x}}_l^T, x_m, \mathbf{\hat{x}}_r^T]^T$ \\
				\nid \textbf{{Output}:} $\mathbf{\hat{x}}$ \\
				\hline
			\end{tabular}\vspace{-0.0cm}
		\end{center}
	\end{table}
\end{center}
\vspace{-0.3cm}
\noindent where the vector $\mathbf{1}$ is defined as $\mathbf{1}\in {1}^{K\times1}$ and the quantizer $Q : \mathbb{R} \mapsto \mathcal{N} $,  $z_1 = Q(y, -K, K)$ is a mapping of $y \in \mathbb{R}$ to the constellation of $\{\pm K, \pm (K-2), ...\}$. Furthermore, let $m_1$, $m_2$, $m_3$, $m_{11}$, $k_1$, $k_2$ and $k_3$ represent the number of $-1$'s at $(1,2)$, $3$, $4$, $1$, $6$, $7$, $8$ locations of $\mathbf{\hat{x}}$, respectively. Note that when $z_1 = K$ or $z_1 = -K$ only one comparison is required. The algorithm proceeds by computing $n$, $n_l$ and $n_r$, which denote the number of $-1$'s in $\mathbf{\hat{x}}$, $[\hat{x}_1,\dots, \hat{x}_{(K-1)/2} ]$ and $[\hat{x}_{(K-1)/2+1},\dots, \hat{x}_K]$, respectively. 

\vspace{-0.3cm}
\begin{center}
	\begin{table}[h]
		\begin{center}
			\begin{tabular}{l}
				\hline \hline \rule{0pt}{3ex} 
				\nid \textbf{SubDecoder Algorithm}  \\
				\hline \rule{0pt}{3ex} 
				\nid \textbf{{Input}:} $\mathbf{y}$, $n$, $n_l$, $n_r$ \\
				\hspace{0.3cm} 1: \!\! \textbf{If}\! $n_l = 0$, $[m_1,m_2,m_3,m_{11}] \gets [0,0,0,0]$, $S_l \gets 1$\\ 
				\hspace{0.3cm} 2:  \textbf{elseIf} $n_l = 4$,\!\!  $[m_1,m_2,m_3,m_{11}]\!\! \gets \!\! [2,1,1,1]$,\!\! $S_l \!\!\gets 1$\\ 
				\hspace{0.3cm} 3: \textbf{If} $n_r = 0$, $[k_1,k_2,k_3] \gets [0,0,0]$, $S_r \gets 1$ \\ 
				\hspace{0.3cm} 4: \textbf{elseIf} $n_r = 3$,  $[k_1,k_2,k_3] \gets [1,1,1]$, $S_r \gets 1$\\ 
				\hspace{0.3cm} 5: \textbf{If} $S_l = 1$ AND $ S_r = 0$, \\
				\hspace{0.3cm} 6:  \hspace{0.3cm}  $[k_1,k_2,k_3] \!\!\! \gets \!\!rightDecoder(\mathbf{y}, m_1, m_2, m_3, m_{11}) $ \\
				\hspace{0.3cm} 7: \textbf{If} $S_l = 0$ AND $ S_r = 1$, \\
				\hspace{0.3cm} 8:  \hspace{0.3cm}  $[m_1,m_2,m_3,m_{11}] \!\!\gets \!\! leftDecoder(\mathbf{y}, k_1, k_2, k_3) $ \\
				\hspace{0.3cm} 9: \textbf{else}, $S_l = 0$ AND $ S_r = 0$ \\
				\hspace{0.3cm}10:  \hspace{0.3cm}  $[m_1,m_2,m_3,m_{11},k_1,k_2,k_3] \gets lrDecoder(\mathbf{y})$ \\
				\hspace{0.3cm}11:  $[\hat{x}_1,\hat{x}_2,\hat{x}_3, \hat{x}_4]\!\! \gets\!\! -2[m_{11}, (m_1\!-\!m_{11}),m_3, m_2] + 1$  \\
				\hspace{0.3cm}12:  $\hat{x}_5 \gets -2(n-n_l - n_r) + 1$  \\
				\hspace{0.3cm}13: $[\hat{x}_6,\hat{x}_7,\hat{x}_8] \gets -2[k_1, k_2, k_3] + 1$   \\
				\nid \textbf{{Output}:} $\mathbf{\hat{x}}$ \\
				\hline
			\end{tabular}\vspace{-0.0cm}
		\end{center}
	\end{table}
\end{center}
\vspace{-0.3cm}
For the case of $\mathbf{C}_{2\times 3}^1$ the decoding is trivial and will not be covered in this article, instead we start with the non-symmetric case of $\mathbf{C}_{4\times 8}^2$. The FDA shown in the table above calls the \textit{subDecoder} at line $8$ with $[y_1,\dots,y_4]^T$, $n_l$ and $n_r$ parameters.
This algorithm will proceed in four different paths depending on $n_l$ and $n_r$. If $n_l$ is $0$ or $4$ then the \textit{leftDecoder} will never be called and will assign $[m_1,m_2,m_3,m_{11}] = [0,0,0,0]$ or $[m_1,m_2,m_3,m_{11}] = [2,1,1,1]$,  respectfully. Similarly, if $n_r$ is $0$ or $3$ then the \textit{rightDecoder} will never be called and will assign $[k_1,k_2,k_3] = [0,0,0]$, or $[k_1,k_2,k_3] = [1,1,1]$, respectfully. Therefore, the trivial case is when both the \textit{leftDecoder} and the \textit{rightDecoder} are not required, other scenarios are the \textit{rightDecoder} is called when the \textit{leftDecoder} is not required, the \textit{leftDecoder} is called when the \textit{rightDecoder} is not required, and the last case is when both left and right decoder, \textit{lrDecoder}, is called.  

\vspace{-0.3cm}
\begin{center}
	\begin{table}[h]
		\begin{center}
			\begin{tabular}{l}
				\hline \hline \rule{0pt}{3ex} 
				\nid \textbf{rightDecoder Algorithm}  \\
				\hline \rule{0pt}{3ex} 
				\nid \textbf{{Input}:} $\mathbf{y}$, $n_r$, $m_1$, $m_2$ \\
				\hspace{0.3cm} 1:  $y_{3m} \gets (y_3 -1)/2 - m_2 +m_1 $\\
				\hspace{0.3cm} 2:  $z_{3m} \gets Q(y_2, -1,+1)$\\
				\hspace{0.3cm} 3:  $k_2 \gets \floor{(z_{3m}+n_r)/2}$\\
				\hspace{0.3cm} 4:  $k_3 \gets z_{3m}+n_r - 2k_2$\\
				\hspace{0.3cm} 5:  $k_1 \gets n_r - k_2 - k_3$  \\
				\nid \textbf{{Output}:} $[k_1,k_2,k_3]$ \\
				\hline
			\end{tabular}\vspace{-0.0cm}
		\end{center}
	\end{table}
\end{center}
\vspace{-0.3cm}
\noindent The \textit{rightDecoder} and the \textit{leftDecoder} decoders are straightforward, having the knowledge of $(\mathbf{y}, n_r, m_1, m_2)$ the \textit{rightDecoder} computes $(k_1,k_2,k_3)$ and similarly, having the knowledge of $(\mathbf{y}, n_l, k_1, k_2)$, the \textit{leftDecoder} computes $(m_1,m_2,m_3, m_{11})$. The last \textit{lrDecoder} computes $(m_1,m_2,m_3, m_{11}, k_1, k_2, k_3)$ given only $(\mathbf{y}, n_l, n_r)$. Note the parameters in the \textit{leftDecoder} and the \textit{lrDecoder} are computed as such; $\delta_{min} = -rnd(3(n_l+1)/5)$, $\delta_{max} = \mod{(rnd(3n_l/5),2)}$, $\beta_{min} = (\sgn(\eta - 1/10)+1)\eta/2$ and $\beta_{max} = \lambda (\zeta - 3)/2- 1 $, where $\eta = \zeta + \delta_{min} - \delta_{max} -1$, $\lambda = \sgn(31/10-\zeta)+1$ and $\zeta$ is the index of the constellation returned by $Q(\cdot)$ function.

\vspace{-0.3cm}
\begin{center}
	\begin{table}[h]
		\begin{center}
			\begin{tabular}{l}
				\hline \hline \rule{0pt}{3ex} 
				\nid \textbf{leftDecoder Algorithm}  \\
				\hline \rule{0pt}{3ex} 
				\nid \textbf{{Input}:} $\mathbf{y}$, $n_l$, $k_1$, $k_2$ \\
				\hspace{0.3cm} 1:   $y_{3k} \gets (y_3 -1)/2 $\\
				\hspace{0.3cm} 2:   $z_{3k} \gets Q(y_2, -k_1+k_2+\delta_{min},-k_1+k_2+\delta_{max})$\\
				\hspace{0.3cm} 3:   $m_2 \gets \floor{(z_{3k}-k_2+k_1+n_l)/2}$\\
				\hspace{0.3cm} 4:   $m_3 \gets z_{3k}-k_2+k_1+n_l - 2m_2$\\
				\hspace{0.3cm} 5:   $m_1 \gets n_l - m_2 - m_3$\\
				\hspace{0.3cm} 6:   \textbf{If} $m_1 = 2$,   $m_{11} \gets 1$ \\
				\hspace{0.3cm} 7:   \textbf{elseIf} $m_1 = 0$,  $m_{11} \gets 0$ \\
				\hspace{0.3cm} 8:   \textbf{elseIf} $y_4/2 - k_1 - m_2+k_2\geq -0.5$, $m_{11} \gets 0$ \\
				\hspace{0.3cm} 9:  \textbf{else},   $m_{11} \gets 1$ \\
				
				\nid \textbf{{Output}:} $[m_1,m_2,m_3, m_{11}]$ \\
				\hline
			\end{tabular}\vspace{-0.0cm}
		\end{center}
	\end{table}
\end{center}
\vspace{-0.3cm}
Having all the required information now the \textit{subDecoder} assigns  $[\hat{x}_1,\hat{x}_2,\hat{x}_3, \hat{x}_4,\hat{x}_6,\hat{x}_7,\hat{x}_8] = -2[m_{11}, (m_1-m_{11}),m_3, m_2, k_1, k_2, k_3] + 1$ and $\hat{x}_5 = -2(n-n_l - n_r) + 1$. Now we completed the case when $K=8$, the rest of the FDA proceeds by applying the general \textit{decoder} algorithm with the inputs of $\mathbf{\hat{y}}_l$ and $\mathbf{\hat{y}}_r$ to obtain $\mathbf{\hat{x}}_l$ and $\mathbf{\hat{x}}_r$, respectively, to find the middle element $x_m =z_l -(\mathbf{\hat{x}}_l^T\mathbf{1} +\mathbf{\hat{x}}_r^T\mathbf{1})$. The decoded data is $\mathbf{\hat{x}} = [\mathbf{\hat{x}}_l^T, x_m, \mathbf{\hat{x}}_r^T]^T$. In the following section, we discuss the analytically performance of the proposed fast decoder.

\vspace{-0.3cm}
\begin{center}
	\begin{table}[H]
		\begin{center}
			\vspace{0.3cm}
			\begin{tabular}{l}
				\hline \hline \rule{0pt}{3ex} 
				\nid \textbf{lrDecoder Algorithm}  \\
				\hline \rule{0pt}{3ex} 
				\nid \textbf{{Input}:} $\mathbf{y}$, $n_l$, $n_r$ \\
				\hspace{0.3cm} 1:   $y_{3n} \gets (y_3 -1)/2$, $d_{3} \gets e^{10}$\\
				\hspace{0.3cm} 2:   $z_{3n} \gets Q(y_2, -\delta_{min}-1,\delta_{max}+1)$\\
				\hspace{0.3cm} 3:   \textbf{for} $\delta_3 \in \{-1+\beta_{min},\dots, -1+\beta_{max} \}$ \\
				\hspace{0.3cm} 4:   \hspace{0.3cm} $m_2^{\prime} \gets \floor{(z_{3n}-\delta_3 +n_l)/2}$\\
				\hspace{0.3cm} 5:  \hspace{0.3cm} $m_3^{\prime} \gets z_{3n}-\delta_3 +n_l- 2m_2^{\prime}$\\
				\hspace{0.3cm} 6:  \hspace{0.3cm} $m_1^{\prime} \gets n_l - m_2^{\prime} - m_3^{\prime}$, $k_2^{\prime} \gets \floor{(\delta_3+n_r)/2}$\\
				\hspace{0.3cm} 7:  \hspace{0.3cm} $k_3^{\prime} \gets n_r + \delta_3- 2k_2^{\prime}$, $k_1^{\prime} \gets n_r - k_2^{\prime} - k_3^{\prime}$\\
				\hspace{0.3cm} 8:  \hspace{0.3cm} \textbf{If} $m_1^{\prime} == 2$,  $m_{11}^{\prime} \gets 1$ \\
				\hspace{0.3cm} 9:  \hspace{0.3cm} \textbf{elseIf} $m_1^{\prime} == 0$,  $m_{11}^{\prime} \gets 0$ \\
				\hspace{0.3cm}10:  \hspace{0.3cm} \textbf{elseIf} $y_4/2 - k_1^{\prime} - m_2^{\prime}+k_2^{\prime}\geq -0.5$, $m_{11}^{\prime} \gets 0$  \\
				\hspace{0.3cm}11:   \hspace{0.3cm} \textbf{else}  $m_{11}^{\prime} \gets 1$ \\
				\hspace{0.3cm}12:   \hspace{0.3cm} \textbf{if}  $d_3^{\prime} \gets |y_4/2 + m_{11}^{\prime} - m_2^{\prime} - k_1^{\prime} +k_2^{\prime}| < d_3 $\\
				\hspace{0.3cm}13:  \hspace{0.4cm} $[m_1,m_2,m_3, m_{11}]\! \gets\! [m_1^{\prime},m_2^{\prime},m_3^{\prime}, m_{11}^{\prime}]$\\
				\hspace{0.3cm}14:  \hspace{0.4cm} $[k_1,k_2,k_3]\! \gets\! [ k_1^{\prime},k_2^{\prime},k_3^{\prime}]$\\
				\hspace{0.3cm}15:  \hspace{0.4cm} $d_3 \gets d_3^{\prime}$ \\
				\nid \textbf{{Output}:} $[m_1,m_2,m_3, m_{11},k_1,k_2,k_3]$ \\
				\hline
			\end{tabular}\vspace{-0.0cm}
		\end{center}
	\end{table}
\end{center}
\vspace{-0.3cm}

\section{{Complexity Analysis}}
\label{performanceAnalysis}
The proposed decoder, discussed in Section \ref{decoder}, deciphers all the users data at the receiver side in a recursive manner. In this section, we  demonstrate the computational complexity analytically. It is important to state that the proposed FDA neither requires any multiplications nor additions, instead, only a few comparisons are performed in the $Q(\cdot)$ function. First, we will look at the average number of comparisons required for the $\mathbf{C}_{4\times 8}^2$ case, whose decoding algorithm is presented in the \textit{subDecoder} algorithm. Since, our proposed $\mathbf{C}_{4\times 8}^2$ matrix is non-symmetric, we will analyze the complexity of decoding all the $2^{8}$ possible input vectors. By closely analyzing FDA algorithm the comparison required for $n=0, 1, 2, 3, 4, 5, 6, 7, 8$ are $1, 25, 144, 289, 488, 369, 155, 28, 1$, respectively, and there are $\binom{8}{n}$ of input vectors per $n$. There are a total of $1500$ comparisons, hence, the average computational complexity is $T_2 = \frac{1500}{256}=5.86$ comparisons. The recursive structure of our proposed matrices for $i\geq3$ possess symmetries that enables us to present the general case. In order to express the relationship for $T_i$, where $i\geq3$, we will first introduce a few definitions. Let us define

\vspace{-0.3cm}
\begin{eqnarray}
G_i &=& \!\!\!\!\!\sum_{j=0}^{2^i+2^{(i-3)}-1} \!\!\binom{2^{(i+1)}+2^{(i-2)}-1}{j}(j+1),
\label{formGi}
\end{eqnarray}

\begin{eqnarray}
&H_i& = \sum_{j=1}^{2^i+2^{(i-3)}-1} \{\binom{2^{i}+2^{(i-3)}-1}{\ceil{\frac{j-1}{2}}}^2(j+1) \nonumber\\
&+&\!\!\!\!\!\!\! 2\!\!\!\sum_{k=0}^{\floor{\frac{j-1}{2}}} \!\!\!\!\binom{2^i+2^{(i-3)}-1}{k}\binom{2^i+2^{(i-3)}-1}{j-k}(2k+1) \nonumber\\
&+& \!\!\!\!\!\!\!2\!\!\!\sum_{k=0}^{\floor{\frac{j-2}{2}}} \!\!\!\!\binom{2^i+2^{(i-3)}-1}{k}\binom{2^i+2^{(i-3)}-1}{j-k-1}(2k+2)\}, \nonumber\\
\label{formHi}
\end{eqnarray}

\vspace{-0.5cm}
\begin{eqnarray}
&U_i& = 4(2^{2^i-1}-2) + 2\!\!\!\!\!\!\!\sum_{j=2}^{2^i+2^{(i-3)}-1} \!\!\!\{\binom{2^{i}+2^{(i-3)}-1}{\ceil{\frac{j-1}{2}}}^2 \nonumber \\
&+&\!\!\!\!\!\!\! 2\sum_{k=1}^{\floor{\frac{j-1}{2}}} \binom{2^i+2^{(i-3)}-1}{k}\binom{2^i+2^{(i-3)}-1}{j-k} \nonumber\\
&+&\!\!\!\!\!\!\! 2\sum_{k=1}^{\floor{\frac{j-2}{2}}} \binom{2^i+2^{(i-3)}-1}{k}\binom{2^i+2^{(i-3)}-1}{j-k-1}\},
\label{formUi}
\end{eqnarray}
\noindent where $G_i$ is the number of comparisons that are required in the first call of the $Q(\cdot)$ function. If the input vector contains $j$ number of $-1$'s, in $Q(\cdot)$ function it needs $(j+1)$ comparisons, as shown in (\ref{formGi}).  Note that due to symmetry, we do not consider all the input vectors $\mathbf{x}\in \{\pm 1\}^{K\times1}$, instead, only half of them, i.e., $2^i+2^{(i-3)}-1$. The $H_i$ is related to the number of comparisons required in the second call of the $Q(\cdot)$ function, while the last term $U_i$ shows how many times left and/or right sub-decoders are called. The general relation for $i\geq3$ can be expressed as
\begin{eqnarray}
T_i \!\! \!\! &=& \!\! \!\! \frac{1}{2^{2^{(i+1)}+2^{(i-2)}-2}}\!\!\!\left[G_i + H_i + U_i\times \hat{T}_{i-1}\right],
\label{formTi}
\end{eqnarray}
\noindent where 
\begin{eqnarray}
\hat{T}_{i-1} \!\! \!\! &=& \!\! \!\! \frac{1}{2^{2^{i}+2^{(i-3)}-2}-1}\left[2^{2^{i}+2^{(i-3)}-2} T_{i-1}-G_{i-1} \right], \nonumber
\label{formThati}
\end{eqnarray}
\noindent is the modified $T_{i-1}$ in which the number of comparisons in the first call of the $Q(\cdot)$ calculations are excluded.

In Table \ref{table:complexity}, we show the complexity results for ($4\times8$), ($8\times17$), ($16\times35$) using the proposed FDA and ML algorithms. As we can see, the complexity of ML decoder increases exponentially, while the proposed decoder has fairly small complexity even for a relatively large matrix size ($16\times35$).
\vspace{-0.2cm}
\begin{table}[h]
	\caption{Complexity Of the Proposed Ternary Codes} 
	\centering 
	\begin{tabular}{l c c c c} 
		\hline\hline \rule{0pt}{3ex}  
		\bf{Decoder} & \bf{Complexity} & ($\mathbf{4\times8}$) & ($\mathbf{8\times17}$) & ($\mathbf{16\times35}$) \\ [0.5ex]
		\hline \rule{-3pt}{3ex}  
		Proposed & Comparisons & $5.86$  & $17.98$  & $50.24$\\[1ex]
		ML & Comparisons & $2^8$ & $2^{17}$ & $2^{35}$ \\[1ex]
		\hline 
	\end{tabular}
	\label{table:complexity}
\end{table}
\vspace{-0.2cm}

\section{Simulation results}
\label{simulation}
\begin{center}
	\begin{figure}[h]
		\hspace*{-0.45cm}
		\includegraphics[width=3.55 in]{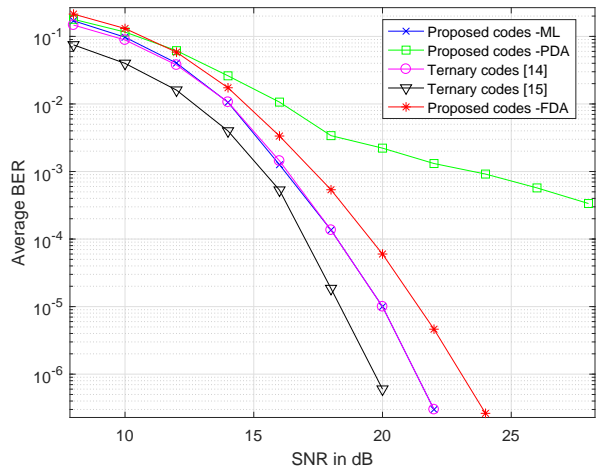}
		\centering \caption{Average BER vs SNR for the UD codes $\mathbf{C}_{4\times 8}^2$.} \label{C2}
	\end{figure}
\end{center}

\begin{center}
	\begin{figure}[h]
		\hspace*{-0.45cm}
		\includegraphics[width=3.55 in]{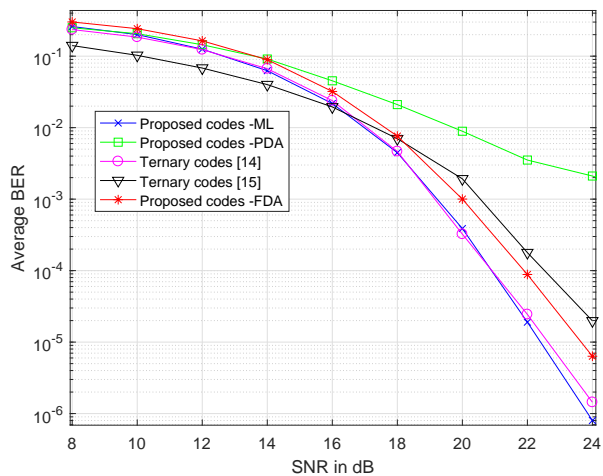}
		\centering \caption{Average BER vs SNR for the UD codes $\mathbf{C}_{8\times 17}^3$.} \label{C3}
	\end{figure}
\end{center}

In this section, we evaluate the performance of the synchronous CDMA over an AWGN channel employing our proposed ternary uniquely decodable codes at the physical layer. All the simulations at the physical layer of the proposed scheme is performed in Matlab. We consider wireless transmission with the number of users $K=8$ and $K=17$. Each user $k$ spreads its data $x_k \in \{\pm 1\}$, using BPSK modulation and the proposed ternary code $\mathbf{c}_k$, and then transmits through an AWGN channel.
At the receiver, MUD is performed using our proposed FDA decoder. For comparison purposes, we compare FDA algorithm with the probabilistic data association (PDA) \cite{Romano01} and the optimum ML decoders. In addition to that in our simulations we have included code constructions from \cite{Marvasti2012} and \cite{Marvasti2016} along with their decoders. Although those presented in \cite{Marvasti2012} and \cite{Marvasti2016} as well as our proposed code sets have the $K < K_{max}^t$, our proposed code sets have larger $K = 2^{(k+1)}+2^{(k-2)}-1$ compared to $K = 2^{(k+1)}-1$ and $K = 2^{(k+1)}-2$, as indicated in the Table \ref{table:ternary}. As an example, for $L= 4, 8, 16, ...$ our code constructions produces $K$, which is larger than the $K$s produced in \cite{Marvasti2012} by $2^0, 2^1, 2^2, ...$ , respectively. In Fig. \ref{C2}, we plot the BER performance averaged over all the different users for our proposed UD code set $\mathbf{C}^2_{4\times 8}$, and we compare them with the $\mathbf{C}_{4\times 7}$ and $\mathbf{C}_{4\times 6}$ constructions presented in \cite{Marvasti2012} and \cite{Marvasti2016}. Specifically, for our proposed UD code set, we perform FDA, PDA and ML decoders, as for the other constructions we used their proposed low-complexity decoders. Similarly, in Fig. \ref{C3}, we plot the BER performance averaged over all the different users for our proposed UD code set $\mathbf{C}^2_{8\times 17}$, and we compare them with the $\mathbf{C}_{4\times 7}$ and $\mathbf{C}_{8\times 14}$ constructions presented in \cite{Marvasti2012} and \cite{Marvasti2016}. There is a trade-off between the number of users, $K$, and BER performance, however, we can observe from Figs. \ref{C2} and \ref{C3} that our propose UD code set performance is as good as the code constructions in \cite{Marvasti2012}. For a BER of $10^{-3}$ the performance of FDA is about $1$dB worse than the ML decoder. In other words, our proposed FDA achieves near-ML performance without having an exponentially complex algorithm. It is obvious that overloaded UD code sets from Table \ref{table:ternary} can potentially increase the user capacity by more than double when $L$ is large.
\vspace{-1.0cm}

\vspace{-0.0cm}
\section{Conclusion}
\label{conclusion}
In this paper, we have introduced new uniquely decodable (UD) ternary code sets for highly overload synchronous code-division multiple-access (CDMA) systems. In comparison to the current state-of-the-art ternary code sets, which have low-complexity decoders, the proposed construction obviously has larger $K < K_{max}^t$. Moreover, using the structure of the proposed code sets, we developed recursive fast decoder algorithm (FDA) that uses only a few comparisons and can allow the users to uniquely recover the information bits at the receiver side. The proposed FDA has much lower computational complexity compared to the maximum likelihood (ML) decoder, which has a high complexity for even moderate code length. Simulation results show that the performance of the proposed decoder is almost as good as the ML decoder in an additive white Gaussian noise (AWGN) channel.



%

\end{document}